\documentclass{article}

\usepackage{amsfonts, amsmath, amssymb, amsbsy}

\usepackage{graphicx}
\usepackage{color} 
\usepackage{hyperref} 

\begin{document}

\begin{center}
\textbf{\LARGE Photometric variability of the nova-like object V380\,Oph in 1976--2016. }
\end{center}

\begin{center}
\textbf{Shugarov S.$^{1,2}$, Golysheva P.$^2$, Sokolovsky K.$^{2,3,4}$, Chochol D.$^{1}$}
\end{center}

\begin{center}
{\it
\noindent $^1$Astronomical Institute of the Slovak Academy of Sciences, Tatransk\'a Lomnica, Slovakia \\
$^2$Sternberg Astronomical Institute, Moscow State University of M.V.Lomonosov, Moscow, Russia \\
$^3$Astro Space Center of Lebedev Physical Institute, Moscow \\
$^4$IAASARS, National Observatory of Athens, Greece}
\end{center}

\begin{abstract}

We combined photographic, photoelectric and CCD observations of the nova-like variable V380\,Oph
to get a light curve spanning the time range of 40 years.
While the typical high-state brightness of V380\,Oph was $R \sim 14.5$, two low-brightness episodes
identified in 1979 ($B_{pg} \sim 17.5$) and 2015 ($R \sim 19$) confirm its classification as
a VY\,Scl-type ``anti-dwarf nova''. The Fourier period analysis of photoelectric and CCD $V$ and $B$
observations obtained in 2002--16 revealed the presence of two periods $0^d.148167$ and $4^d.287$,
that may be associated with negative superhumps and disc precession. We also compared measurements
obtained with the iris micro-photometer and flatbed scanner at the same plates and found an
agreement within the expected accuracy of photographic photometry.
\end{abstract}

\section{Historical overview}
V380\,Oph was discovered as a variable star by C. Hoffmeister (1929). Hope (1938) suggested
that the object belongs to the class of Mira-type stars. Meinunger (1965) examined the photographic
plates of Sonneberg Observatory and concluded that the star might be either an eclipsing binary
or a RR Lyrae-type variable. Bond (1979) first suspected that V380\,Oph
may be a cataclysmic variable (CVs). Shafter (1983, 1985) showed that the spectra of V380\,Oph
exhibited strong Balmer emissions and HeII 4686 \AA~ line. By analysis of radial
velocities he found the orbital period as $\sim0^d.158$. The first detailed photometric
study of V380\,Oph was carried out by Shugarov et al. (2005, 2007). The authors analyzed
$\sim$ 220 photographic (1976 -- 90) and $\sim$  1300 photoelectric and $BVR_C$ CCD
(2002 -- 04) observations and showed that the object was at a low state ($\sim 17.5 pg$) in 1979,
and $\sim 14^m$ at other times. They found two periods of $0^d.14817$ and
$4^d.5135$ in 2002 -- 04 data, probably related to the orbital period and disc precession,
respectively. Rodriguez-Gil et al. (2007) used H$\alpha$ radial velocities to find the orbital
period $0^d.154107$ and classified object as the SW Sex-type system.
The short orbital period of V380\,Oph and detected large depression of
luminosity include it also in VY\,Scl-type nova-like CVs (anti-dwarf novae).
DW\,UMa is another example of a system displaying both VY\,Scl and SW\,Sex
features (Dhillon et al., 2013).

\section{The photographic, photoelectric and CCD photometry.}

During 1976--1995 about 220 photo-plates of the 66\,Oph field, covering the
position of V380\,Oph, were obtained using the 40~cm astrograph of the Southern Station of the Sternberg Astronomical
Institute (SAI) in Crimea. In the past, the V380\,Oph plates were measured by an iris micro-photometer.
Recently, the same plates were digitized with a flatbed scanner (Kolesnikova et al., 2008).
Thus, we have photographic magnitudes, obtained by different methods. Since the plate size is large (30 x 30 cm),
the digitized plate was divided into $1^\circ\times1^\circ$ subfields that were processed independently of each other. 
The magnitude scale was calibrated using APASS $B$ magnitudes of all UCAC4  stars in the subfield
(Sokolovsky et al., 2014). V380\,Oph, visible in two overlapping subfields, was measured twice
on each plate using two different sets of comparison stars. Fig.~1 (left) shows the measured magnitude
of the variable at the first section of the plate versus its magnitude at the second section
of the same plate. There is a very good correlation between the two independent measurements
of the variable star. The standard deviation (SD) of $\sim 0^m.05$ reflects the calibration
uncertainty associated with the choice of a specific set comparison stars.
We used the average of these two magnitude measurements obtained for each plate.
Fig.~1 (right) shows the magnitude $B_{scan}$, obtained by scanning, with respect to the magnitude
$B_{pg}$, measured by the micro-photometer. There is a small systematic discrepancy of magnitudes
and the slope of the curve differs from $45^\circ$. We applied the transformation equation
$B_{pg}=1.10\cdot B_{scan}-1.725$ and averaged the magnitudes measured by micro-photometer and scanning methods
to construct the final photographic light curve. The SD of $0^m.13$ is close to the expected measurement accuracy 
on photo plates. The $B_{pg}$ magnitudes of V380\,Oph (1975-96) are
presented in Fig.~2 together with our photoelectric and CCD photometry obtained at the 60-125\,cm telescopes
in Russia and Slovakia (2002-16). A deep fading of brightness up to $4^m$ was detected in 2015.
At the same time the optical companion $\sim 21^m$ located southwest from V380\,Oph at an angular
distance $\sim 5^{\prime\prime}$ was seen.

\begin{figure}[!]
\centerline{\includegraphics[width=0.44\textwidth,trim=0cm 0.5cm 0.4cm 0.4cm,clip]{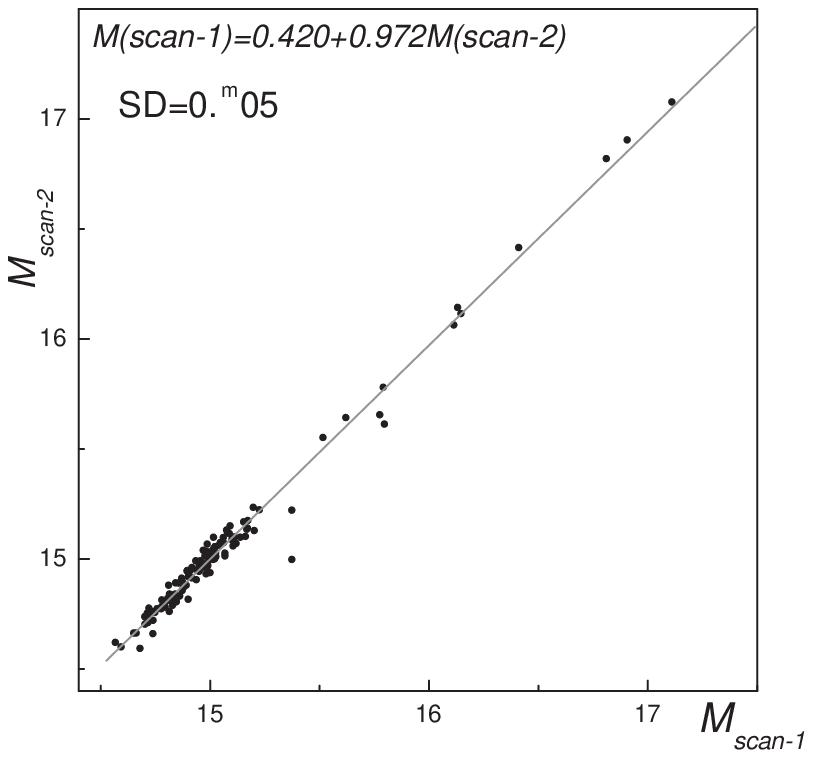}
            \hfill
            \includegraphics[width=0.42\textwidth,trim=0.2cm 0.48cm 0.1cm 0.6cm,clip]{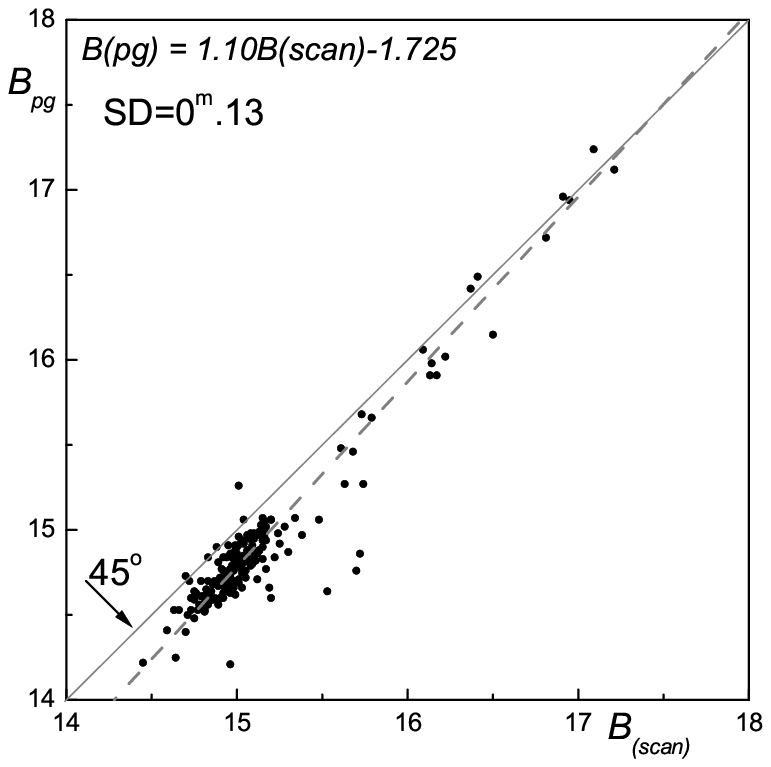}}
\vspace{-2mm}
\caption{Left: Measured magnitude of the variable at the first section (subfield)
of the photographic plate versus its magnitude at the second section of the same plate
for 220 plates taken in 1976--95. Right: The dependence of V380\,Oph magnitude measured by
micro-photometer versus its magnitude obtained by scanning of the photographic plate.}

\end{figure}

\begin{figure*}[!]
\centering
\includegraphics[width=0.97\textwidth,trim=0.3cm 0.4cm 0.4cm 0.6cm,clip]{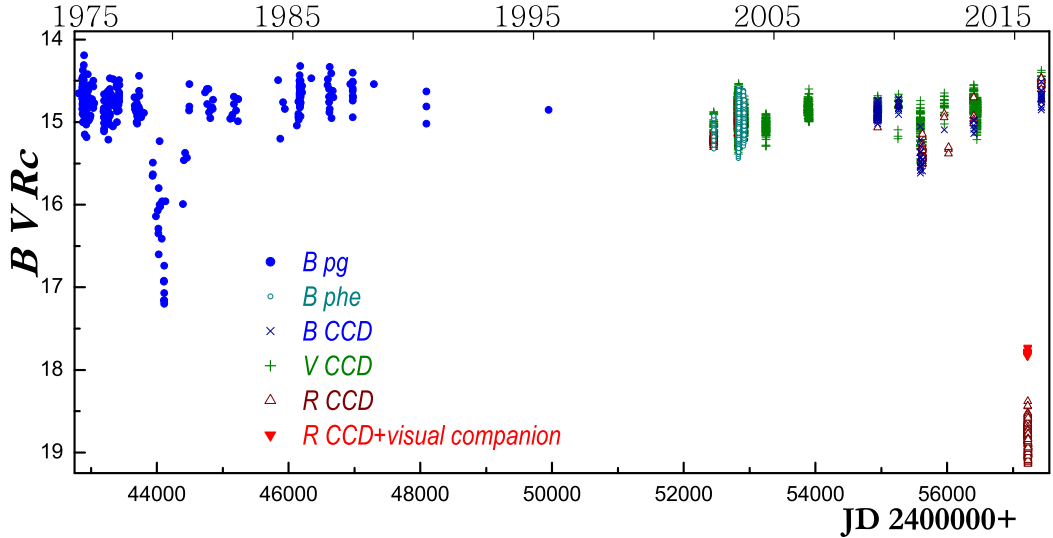}
\vspace{-5mm}
\caption{The light curves in $BVR_C$ bands during the last 40 years. The depressions were detected
in 1979 and 2015.}
\vspace{-2mm}
\label{logo}
\end{figure*}

Photoelectric and CCD observations have obviously a better accuracy and time resolution
than the photographic ones. The period $P_2 = 4^d.287$, found by Fourier period analysis
of $V$ and $B$ 2002--16 data, is 5\% shorter than the value reported by Shugarov et al. (2005).
The $V$ phase light curve is shown in Fig.~3 (left). We subtracted the wave associated with $P_2$
from the $V$ and $B$ data and found by Fourier period analysis of the residuals the best period
$P_1= 0^d.148167$ (Fig.~3, right), in full agreement with the value found by Shugarov et al. (2005)
from 2002--04 observations.

\begin{figure}[h]
\vspace{-2mm}
\centerline{\includegraphics[width=0.48\textwidth,trim=0cm 0.7cm 0.4cm 0.5cm,clip]{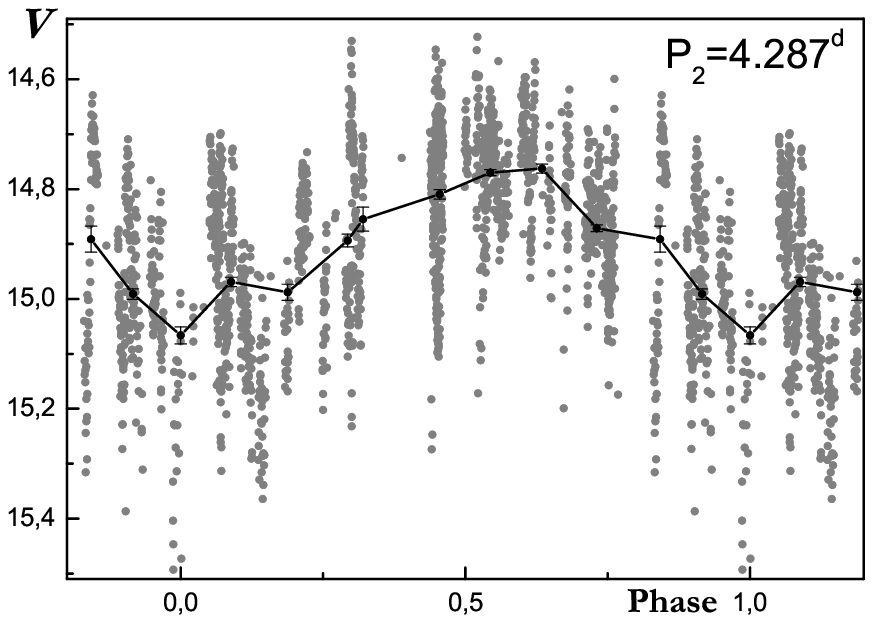}
            \hfill
            \includegraphics[width=0.5\textwidth,trim=0cm 0.7cm 0.4cm 0.5cm,clip]{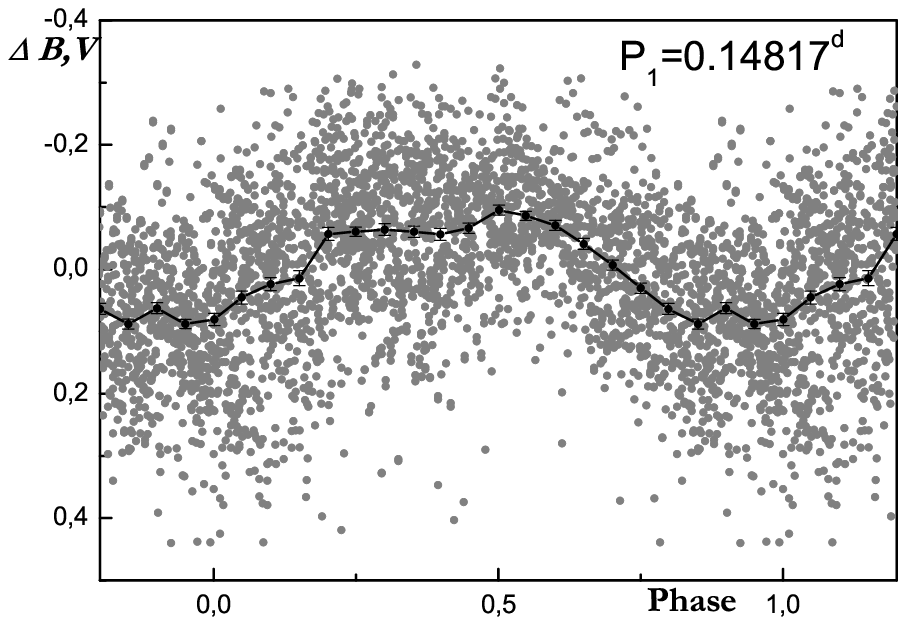}}
\vspace{-2mm}
\caption{Left: The $V$ light curve folded with the period of $4^d.287$.
Right: The phase diagram of the $B,V$ data residuals after removing the $4^d.287$ wave, folded
with the period of $0^d.148167$.}
\vspace{-3mm}
\end{figure}

\section{The joint analysis of archival photographic and modern observations and final remarks}

The Fourier period analysis of photographic data did not reveal the period of $0^d.148$,
although there were some indications of periodicities close to this period with small probability.
The amplitude of the $P_1$ component is about $0^m.15$ (see Fig.~3, right), close to the
limit of precision of photographic measurements. Therefore, 220 photographic observations
obtained in the interval of 20 years are insufficient for the purpose of periods search.
In addition, there is strong flickering in V380\,Oph that further complicates periodicity searches.

Photographic observations are still relevant and important to study
long-term high-amplitude brightness variations. Through these observations we found the
low state of V380\,Oph in 1979 and suggested the existence of a short period.
More precise photoelectric and CCD photometry allowed us to
find the period close to the orbital one, which can be interpreted as the period of negative
superhumps (Rodriguez-Gil et al., 2007).

This article continues the study of CVs and related objects by our group
using the archive negatives of the SAI and other observatories.
Earlier we obtained detailed light curves of classical, symbiotic and X-ray
novae: V1500\,Cyg (Harevich et al., 1975), V616\,Mon (Shugarov, 1976), HR\,Del (Shugarov, 1967),
HM\,Sge (Dokuchaeva, 1977; Chochol et al., 2004), V1680\,Aql (Antipin et al., 2005),
FG\,Ser (Shugarov, et al., 2014), V838\,Mon and V445\,Pup (Goranskij at al., 2004, 2007, 2010),
V718\,Per (Goranskij et al., 1996), Q\,Cyg (Shugarov, 1983), V404\,Cyg (Osminkina et al., 1990).
The negatives of SAI archive allowed Kurochkin (1972) to discover the optical period
of X-ray source HZ\,Her. 
We found or investigated the eclipses of CVs:
AC\,Cnc (Shugarov, 1984; Baidak \& Shugarov, 1986), IP\,Peg (Goranskij et al., 1985), UU\,Aqr
(Volkov et al., 1986), BE~UMa (Kurochkin \& Shugarov, 1992), AY\,Psc (Shugarov, 1984; 
Kazennova \& Shugarov, 1992), DW\,UMa (Kazennova \& Shugarov, 1992).
We detected and studied optical variability of CVs and related objects:
V795\,Her (Mironov et al, 1985), V361\,Lyr and V363\,Lyr (Galkina \& Shugarov,\,1985),
AN\,UMa (Shugarov, 1975), BQ\,Cam (Goranskij, 2001), V1341\,Cyg and V818\,Sco (Basko et al., 1976),
EF\,Peg (Tsesevich et al., 1979), 2MASS\,J01333949+3045405 (Sharov et al, 1997), 
QR\,And (Katysheva \& Shugarov, 2003), DV\,Dra (Pavlov \& Shugarov, 1985) and many other stars.

\vspace{2mm}
{\bf Acknowledgments.} This work was supported by the VEGA grant No.\,2/0002/13
and partially by RFBR grants No\,14-02-00825, 15-02-06178. P.\,Golysheva is grateful to SAIA\,(2015)\,grant.

\end{document}